\documentclass[preprint,12pt,authoryear]{elsarticle}
\usepackage[utf8]{inputenc}
\usepackage[T1]{fontenc}
\usepackage[british,UKenglish,USenglish,english,american]{babel}

\usepackage{amssymb, color}

\usepackage{amsthm}

\newtheorem{lemma}{Lemma}
\newtheorem{definition}{Definition}

\journal{International Journal of Forecasting}

\begin{document}

\begin{frontmatter}



\title{ Extreme events evaluation using CRPS distributions }
%
%

\author[a1,a5]{Maxime Taillardat\corref{correspondingauthor}}
\ead{maxime.taillardat@meteo.fr}
\author[a2]{Anne-Laure Fougères}

\author[a3]{Philippe Naveau}

\author[a4]{Raphaël de Fondeville}
%



\address[a1]{CNRM, Université de Toulouse, Météo-France, CNRS, Toulouse, France.}
\address[a5]{Météo-France, Toulouse, France}
\address[a2]{Univ. Lyon, Université Claude Bernard Lyon 1, CNRS UMR 5208,
Institut Camille Jordan, F-69622 Villeurbanne, France}
\address[a3]{Laboratoire des Sciences du Climat et de l'Environnement, UMR 8212, CEA-CNRS-UVSQ, IPSL \& U Paris-Saclay, Gif-sur-Yvette, France}
\address[a4]{Swiss Data Science Center, ETH Zürich and EPFL, Switzerland}

\cortext[correspondingauthor]{Corresponding author}

\begin{abstract}

Verification of probabilistic forecasts for extreme events has been a very active field of research, stirred by media and public opinions who naturally focus their attention on extreme events, and easily draw biased conclusions.
In this context, classical verification methodologies tailored for extreme events, such as thresholded and weighted scoring rules, have undesirable properties that cannot be mitigated; the well-known  Continuous Ranked Probability Score (CRPS) makes no exception.

In this paper, we define a formal framework to assess the behavior of forecast evaluation procedures with respect to extreme events, that we use to point out that assessment based on the expectation of a proper score is not suitable for extremes.
As an alternative, we propose to study the properties of the CRPS as a random variable using extreme value theory to address extreme events verification.
To compare calibrated forecasts, an index is introduced that  summarizes the ability of probabilistic forecasts to predict extremes. Its strengths and limitations are discussed using both theoretical arguments and simulations.

\end{abstract}

\end{frontmatter}



 
\section{Introduction}

By definition, the rarity of extreme events makes difficult to issue relevant forecasts, whose performance assessment is an even greater challenge. In particular, the scarcity of extremes imposes that verification   schemes have to be built and understood in a probabilistic sense. 
The general framework for probabilistic forecast evaluation compares an observation $y$  with a probabilistic forecast  $F$, represented by its cumulative distribution function (cdf). The framework also assumes that $y$ is drawn from  a random variable $Y$ with cdf $G$. 
For a better utilization of the forecasts, it is generally convenient, and even recommended \citep{ferro2011extremal}, to further assume that the forecast $F$ is calibrated \citep{dawid1984present, diebold1997evaluating}, i.e., that the predictive distribution resembles the distribution of the observations given the information contained in the  forecast. For a formal definition of auto-calibration (calibration in the following), we refer to the works of \citet{tsyplakov2011evaluating} and \citet{strahl2017cross} summarized in \ref{framework-calib}.

Calibrated forecasts can be commonly evaluated based on their sharpness, also called refinement by \cite{winkler1996scoring}, 
which usually refers to their spread.
This leads to 
the paradigm  of `maximizing sharpness subject to calibration', introduced by \citet{gneiting2007probabilistic} and later formally justified by  \citet{tsyplakov2011evaluating}.

Probabilistic forecasting has become more and more popular over the last years in various fields such as economics and finance \citep{galbraith2012assessing}, demography and social science \citep{RAFTERY2021}, health \citep{henzi2021probabilistic}, energy \citep{hong2016probabilistic}, hydrology and hydraulics \citep{tiberi2021strategies}.
In this work, we focus on weather probabilistic forecasts \citep{leutbecher2008ensemble}.
Indeed, probabilistic forecasts are nowadays issued by most National Weather Services (NWS) and $F$ is known through 
a sample of finite size called ``ensemble'' \cite[see, e.g.,][]{Zamo17}. 
In this context, forecast verification is performed by computing scoring rules 
such as the Continuous Ranked Probability Score (CRPS)
\citep[][]{epstein1969scoring,hersbach2000decomposition,brocker2012evaluating} 
\begin{eqnarray}
 \mathrm{CRPS}(F,y)&=&\int_{-\infty}^{\infty} (F(x) - \mathbf{1}\{x\geq y\})^2 \,dx, \nonumber \\ 
 &=&\mathbb{E}_F|X-y|-\frac{1}{2}\mathbb{E}_F|X-X'|, \label{eq:crps2}
\end{eqnarray}
where $y \in \mathbb{R}$, and $X$ and $X'$ are independent random  variables with  common cdf $F$. %
The CRPS is attractive as it does not require predictive densities, is inferred non-parametrically, and has simple interpretation. 
The right hand side of Equation  (\ref{eq:crps2}) decomposes the CRPS into, in this order, a calibration and a sharpness term \citep{gneiting2007strictly}. Alternative decompositions are  also available; see
 \citet{taillardat2016calibrated, Bessac21}
 and \ref{app:CRPSequivalences}. 

For the forecast evaluation of extreme events, proper weighted scoring rules  were  introduced by 
\citet{gneiting2011comparing}  and  \citet{diks2011likelihood}.
For a non-negative  function  $w(x)$,  the weighted CRPS
\begin{eqnarray}
 \mathrm{wCRPS}(F,y)&=&\int_{-\infty}^{\infty} (F(x) - \mathbf{1}\{x\geq y\})^2 w(x)\,dx, \label{eq:def-wCRPS}\\
 &=&\mathbb{E}_F|W(X)-W(y)|-\frac{1}{2}\mathbb{E}_F|W(X)-W(X')|, \nonumber
\end{eqnarray}
with $W(x)=\int_{-\infty}^x w(t)dt$, aims to emphasize a  region of interest, for instance distributional tails. 
When $w$ is continuous, an alternative expression of the weighted CRPS is available and can be found in \ref{app:CRPSequivalences}.
The choice of the weight function $w(x)$ is complex and depends on the different stakeholders, such as forecast users and forecasters;
see, e.g., \citet[][]{ehm2016quantiles,gneiting2011comparing,patton15,smith2015towards,taillardat2021skewed}.
Even in the hypothetical   case where  $w(x)$ could  be objectively defined, it is essential that the verification process has to be made on the whole set of observations \citep{lerch2017forecaster} and one can wonder if the corresponding  weighted CRPS  correctly discriminates between two competitive forecasts with respect to extreme events. 

In this work, we show that the expected weighted CRPS cannot discriminate forecasts with different extremal tail behaviors, a potentially redhibitory defect for extremal evaluation. 
To address this issue, we view  the CRPS as a random variable.
Its tail behavior  is derived and compared to the tail regime of observations using Extreme Value Theory (EVT) \cite[see, e.g.][]{de2007extreme}.

This work is organized as follows: Section~\ref{taileq} provides an analysis of  the weighted CRPS 
with respect to the notion of tail equivalence, the main backbone of EVT. In particular, we propose a benchmark to compare the tail properties of forecast verification tools allowing us to pinpoint the shortcomings of 
 the CRPS and its weighted counterpart for scoring extreme events. 
In Section~\ref{CRPSasRV}, we study   the CRPS as a random variable and we make theoretical links between  its tail behavior  and the observational tail distribution. 
These mathematical connections help us to  propose and study a new   index to assess the  skill of calibrated probabilistic forecasts with respect to extreme events. The paths and pitfalls of this index and potential future works are discussed in the Section~\ref{blah}.

\section{Limitations of the (w)CRPS as a proper scoring rule for extremes}\label{taileq}
 
\subsection{Tail modelling using EVT}
Thanks to the pioneering work of \citet{gumbel1935valeurs} and \cite{haan1970regular}, EVT provides a theoretically justified framework to model the tail of random variables, more precisely excesses above a large threshold; see, e.g., \cite{embrechts1997modelling, beirlant:2004}.
For any random variable $X$ with cdf $F$, EVT models assume the existence of a domain of attraction, i.e., that there exists a positive auxiliary function $b$, such that 
\begin{equation}\label{eq: domain attraction}
\frac{\overline{F}\{u+ x b(u)\}}{\overline{F}(u)} \longrightarrow \overline{H}(x) > 0 ,\quad u \rightarrow x_F,
\end{equation}
where $\overline{F}=1-F$ corresponds to the survival, also called tail function, and  $x_F = \sup \{x : F(x) < 1\}$ is the upper endpoint of $F$.
Under condition~(\ref{eq: domain attraction}), noted $F \in \mathcal{D}(H)$, the  Pickands-Balkema-de Haan's theorem  \citep{haan1970regular,pickands1975statistical} establishes that $H$ has to belong to the family of generalized Pareto (GP) survival functions, i.e., 
$$
\overline{H}_{\gamma}(x)= \left(1+ \gamma x\right)^{-\frac{1}{\gamma}},
$$
where $x \in \{x : 1+ \gamma x >0\}$.
As a consequence, the GP tail appears to be the ideal candidate to  approximate  the survival function of exceedances over a large threshold $u >0$, i.e.,
$$
\mathbb{P}(X-u\geq x | X > u)  \approx \overline{H}_{\gamma}(x/\sigma)= \left(1+\frac{\gamma x}{\sigma}\right)^{-\frac{1}{\gamma}},
$$
where $x\in \{x : 1+ \gamma x / \sigma>0\}$ and $\sigma>0$.
The GP family   covers the three possible regimes of tail decay which is determined by the value of its tail index~$\gamma$: when $\gamma \neq 0$ the decay is polynomial and has an upper bound when $\gamma <0$.
For $\gamma=0$, the GP survival function becomes exponential, i.e.,  $\overline{H}_{0}(z)=e^{-z/\sigma}$.

\subsection{Tail equivalence and  proper scoring rules} \label{sec:tailequiv}

The comparison of the tail behavior of two random variables, or equivalently their respective cdfs $F$ and $G$, can be framed using the notion of tail equivalence.
\begin{definition}
\cite[][Section 3.3]{embrechts1997modelling} Two random variables $X$ and $Y$ with respective cdf $F$ and $G$ are \textit{tail equivalent} if they have equal upper endpoint $x_F=x_G=x_*$ 
and if their survival functions $\overline{F}$ and $\overline{G}$ satisfy 
$$
\lim_{x\rightarrow x_*} \frac{\overline{F}(x)}{\overline{G}(x)} =c \in (0,+\infty).
$$
\end{definition}
Tail equivalence can also be simply expressed as the equality of tail indexes.
In terms of extremal forecast, we expect that, between two forecasters,  one should favor the one that is tail equivalent to the observations. 
In practice, this may be difficult.  
For instance, consider two GP distributed random variables $X_1$ and $X_2$ with survival functions $\overline{H}_1(x)$ and $\overline{H}_{1 + \epsilon}(x / \sigma)$ with $\sigma= (1 + \epsilon)/(2^{1 + \epsilon}-1)$. By construction, the medians of $X_1$ and $X_2$ are both equal to one. 
Still, their tail behavior widely differ even for small $\epsilon$:
The 100 year return level for $X_1$ is 99, while it is equal to 138 for $X_2$ with  $\epsilon=0.1$. 
In other words,  
if the precedent random variables were to represent water levels,   a small difference of $0.1$ in tail index, implied a difference of $39$ meters which would most likely cause massive and destructive flooding.

This short example illustrates how issuing forecasts with the right tail regime, i.e., as close as possible to the observational one, is a priority for extreme events and that a verification methodology should reward forecast with close, if not equal, tail regime.
Ideally, the measure of forecast performance should give not only the distance but also the `direction', i.e., if the forecast is more likely to over- or under-estimate the high quantiles.
Indeed, let $\gamma_G \in \mathbb{R}$ be the tail index of observations.
If the forecast satisfies $\gamma_F > \gamma_G$, the forecast over-estimates the risk producing a pessimistic or risk averse scenario.
On the contrary, $\gamma_F < \gamma_G$ falls on the optimistic side by under-estimating the likelihood of extreme events.

Classical methods for forecast evaluation, even when designed to focus on extreme events, do not conserve tail equivalence.
For instance, for any positive  $\eta$ and observation distribution $G$, 
%
%
it is always possible to construct a non-tail equivalent cdf $F$, such that
\begin{equation}\label{eq: tail equiv} 
\left|\mathbb{E}_{G}(\mathrm{wCRPS}(G,Y))-\mathbb{E}_{G}(\mathrm{wCRPS}(F,Y))\right| \leq \eta,
\end{equation}
proof can be found in \ref{app:ineq-wCRPS}.
More precisely if $G \in \mathcal{D}(H_{\gamma_G})$, then it is possible  for any arbitrary $\gamma_F \in \mathbb{R}$ to find $F \in \mathcal{D}(H_{\gamma_F})$ satisfying Equation (\ref{eq: tail equiv}).
Thus the CRPS is unable to discriminate properly forecasts with different tail regime, as non-tail equivalent forecasts can perform almost equally well as the ideal forecast $G$.
A detailed illustration of this result for GP forecasts is given in \ref{app:ex-2.2}. We also refer to \cite{brehmerStrokorb19}, who obtained a more general result, proving that proper scoring rule expectations are not suitable to distinguish tail properties, see their Theorem 5.4.


 
\subsection{A benchmark for assessing forecasts of extremes}\label{sec:GEmodel}

Following \citet{gneiting2007probabilistic} and   \citet{strahl2017cross}, we propose a benchmark to assess the behavior of forecast evaluation procedures with respect to tail regimes.
The design relies on a hierarchical model based on Gamma--exponential mixtures 
with $\gamma > 0$ 
\begin{equation}\label{eq: gamma-mixture}
\left\{
\begin{array}{rl}
\Delta & \stackrel{d}{ =  } \Gamma(\gamma^{-1},\gamma^{-1}) \\
Y & \stackrel{d}{ =  } \textrm{Exp}(\Delta) \stackrel{d}{ =  } \textrm{GP}(1,\gamma),
\end{array}
\right.
\end{equation}
where $\textrm{Exp}(\delta)$ refers to an exponential random variable with scale $\delta > 0$. 
The fact that $Y$ follows a heavy tailed GP distribution, see relation (\ref{eq: gamma-mixture}), can be  proved using Laplace transforms.
For  analogy with weather forecasting, we present the benchmark in a temporal setting.
At each time $t = 1,\dots, T > 1$, an observation $y$ is drawn  independently from an exponential distribution whose scale $\delta$ is a realization of $\Delta$.
In this setting, $Y$ has an exponential tail which is conditioned by the information brought by its scale $\delta$, representing the \textit{a priori} knowledge of the system, for instance the weather at previous time. 
Thus the ideal forecast for each time step is $\textrm{Exp}(\delta)$, and requires the knowledge of $\delta$.
Using relation (\ref{eq: gamma-mixture}), we see that the \textit{climatological} forecaster $F_{\rm clim}$ is a GP distribution with tail index $\gamma$ and unit scale.
Climatology is a commonly used forecast reference in meteorology. In other fields, it can be viewed as the unconditional distribution of the truth, and an estimation of a climatological forecast can be done based on a sample of past and analogs observations.
This setting is attractive as the ideal and the climatological forecasters belong to two different regimes of tail decay.

We introduce alternative competitors modelling partial knowledge of the conditional state: the \textit{$\lambda$-informed} forecaster $F_\lambda$, $\lambda \in [0,1]$ is a  mixture between  the climatological and  ideal forecasts, 
 where a weight, say $\lambda \in [0;1]$, indicates the contribution of each one, see Table \ref{GEforecasters} for the definition.


Finally, the  \textit{extremist} forecaster $F_{\rm extr}$
simply adds a multiplicative bias to the ideal forecaster: while it is not calibrated, such forecast has the same tail behavior as the ideal forecaster 
; see \ref{framework-calib} for detailed discussion on calibration.
The benchmark is summarized in Table~\ref{GEforecasters} and later referred to as the  ``Model GE''.
%

\begin{table*}[!h]
\caption{Benchmark to assess the behavior of forecast evaluation procedure with respect to different tail regimes. All forecasts but $F_{\rm extr}$ are calibrated.}
\label{GEforecasters}
\begin{center}
\begin{tabular}{cc}
\hline
\hline
 Forecasts ~$\backslash$  Truth & $Y \stackrel{d}{ =  }  \mathcal{E}\rm{xp}(\Delta)$ where $\Delta \stackrel{d}{ =  } \Gamma(1/\gamma,1/\gamma)$, $1 > \gamma > 0$\\
 \hline
 Ideal $F_{\rm ideal}$ & $\mathcal{E}\rm{xp}(\Delta)$ \\
 Climatological $F_{\rm clim}$  & $\mbox{\rm GP}(1, \gamma)$ \\
 $\lambda$-Informed $F_\lambda$ &  $\lambda \mathcal{E}\rm{xp}(\Delta) + (1-\lambda) \mbox{\rm GP}(1, \gamma)$ \\
Extremist $F_{\rm extr}$ &  $\mathcal{E}\rm{xp}(\Delta/\nu)$, $\nu > 1$\\
\hline
\end{tabular}
\end{center}
\end{table*}

Closed forms of the CRPS are available for each forecast of the proposed benchmark.
For instance, 
the extremist forecast $F_{extr}$, satisfies
\begin{equation}\label{eq:crps-G-Delta}
CRPS(F_{\rm extr},y)=y+\frac{2\nu}{\delta}\exp\left(-\frac{\delta y}{\nu}\right)-\frac{3\nu}{2\delta} \;; 
\end{equation}
Besides, combining (\ref{eq:wCRPS}) and (\ref{eq:crps-G-Delta}) yields the following formula for the  $\lambda$-informed forecast, $\lambda \in [0,1],$  
\begin{eqnarray*}\label{eq:crps-G-lambda}
 CRPS(F_\lambda,y)&=&y+\frac{\lambda^2}{2\delta}+ \frac{2\lambda}{\delta}\left\{\exp(-\delta y)-1\right\} -\frac{2(1-\lambda)}{1-\gamma}\left\{ 1-(1+\gamma y)^{\frac{\gamma-1}{\gamma}}\right\} \\ 
 && \hspace{-2cm} +\frac{2(1-\lambda)^2}{2-\gamma} + \frac{2\lambda(1-\lambda)\gamma^{\frac{-1}{\gamma}}}{\delta^{\frac{\gamma-1}{\gamma}}}\left\{\exp\left(\frac{\delta}{\gamma}\right) {\rm I}\!\Gamma\left(\frac{\gamma-1}{\gamma},\frac{\delta}{\gamma}\right)\right\} \; ,
\end{eqnarray*}
where ${\rm I}\!\Gamma(s,x)= \int_x^{+\infty} e^{-t}t^{s-1}\, dt$. 
Table \ref{crpsmean} gives the relative ratio of the empirical means of the CRPS for the benchmark with $\gamma = 1/4$.
\begin{table*}[!h]
\caption{Relative ratio of the mean CRPS, in percent, with respect to the ideal forecast for the model GE with $\gamma = 1/4$, based on $T=10^6$ observation/forecast pairs.
}
\label{crpsmean}
\begin{center}
\begin{tabular}{cc}

\hline
\hline
   Truth & $Y \stackrel{d}{ =  }  \mathcal{E}\rm{xp}(\Delta)$ where $\Delta \stackrel{d}{ =  } \Gamma(4,4)$\\
 \hline
 Forecasts   &  $\%$ w.r.t. Ideal  \\
 Ideal $F_{\rm ideal}$ & $100\%$\\
   Extremist $\nu=1.1$ & $100.48\%$\\
  0.75-Informed $F_{0.75}$ & $100.90\%$ \\
 0.5-Informed $F_{0.5}$ & $103.58\%$ \\
  Extremist $\nu=1.4$ & $106.68\%$\\
 0.25-Informed $F_{0.25}$ & $108.06\%$ \\
Climatological $F_{\rm clim}$ & $114.33\%$\\
  Extremist $\nu=1.8$ & $122.89\%$\\

\hline
\end{tabular}
\end{center}
\end{table*}
The CRPS being a proper score, the ideal forecast cannot be beaten in average in the Table \ref{crpsmean}. Moreover, there is a clear ranking among calibrated forecasts, based on the nested information sets \citep{holzmann2014role}. Following the principle of tail equivalence presented in Section \ref{sec:tailequiv}, the extremist forecast should be the forecast the closest to the ideal as they both belong to the same regime of tail decay; however, we observe that the CRPS average gives a performance in between the least informed forecaster and the climatology.
An alternative measure for forecast evaluation, satisfying the tail equivalence principle is thus required.
A good candidate commonly used in forecast science is the ROC curve \citep{gneiting2018receiver}.
However, in the case of Model GE, all the ROC curves, except the climatological one, coincide whatever the event,
which illustrates its invariance under calibration \citep{kharin2003roc}.
Further alternatives should thus be investigated.


\section{The CRPS as a random variable}\label{CRPSasRV}

\subsection{The random CRPS and its properties}\label{sec:calibinfo}
Section~\ref{taileq} pointed out the difficulty of summarizing forecast performance for meaningful comparisons for extreme observations.
We illustrated in particular that a single number such as the mean of the CRPS, or its weighted counterpart, fails to deliver relevant comparisons.
As an alternative, we propose to study the distribution of the CRPS when treated as a random variable, see also  \cite{https://doi.org/10.1002/qj.3115,Bessac21}.

For simplicity, we use the setting and corresponding notations of the benchmark presented in Section \ref{sec:GEmodel}. 
From equations (\ref{eq:wCRPS}) and (\ref{eq:crps-G-Delta}), the climatological and ideal scores can be treated as random variables whenever $y_t$ is replaced by $Y_t$. 
At this stage, it is important to remind that a forecast is issue with only a partial knowledge of the system: the exact value of $\delta_t$ and the distribution of $Y_t$ are unknown, and only the observation $y_t$ is available. 
Table~\ref{avail} summarizes quantities that are available to forecasters. 
Thus, to evaluate forecasts performance, it is only possible to compute $\mathrm{CRPS}(F_t,y_t)$ for each $t$.
The climatological distribution, that we now note $G$ 
and whose existence needs to be hypothesised in practice, is characterized by the observed sample $(y_1,\dots, y_t)$, considered as a sample of independent realizations of the random variable $Y$.

 
For any set of forecasts $\{F_t\}_{t = 1,\dots,T}$ and sample $y_1,\dots, y_T$, two types of sets of random variables can be defined:
\begin{equation}\label{eq: nabla}
{\cal S}(F_T)= \{\mathrm{CRPS}(F_t,Y_t)\}_{t=1,\dots, T}  ~~~{\rm and}~~~~   {\cal S}^*(F_T)=\{\mathrm{CRPS}(F_t,Y_{\pi(t)})\}_{t=1,\dots, T}\; ,
\end{equation}
where $\pi$ is a random permutation of $\{1,\dots,n\}$.
Applying $\pi$ breaks the conditional dependence between $y_t$ and $F_t$, quantified by $\delta_t$ in the benchmark, creating alternative less informative forecasts. 
Thus for a given forecaster, represented by the set $F_T = \{F_t\}_{i = 1,\dots,T}$ and permutation $\pi$, we introduce two random variables ${\cal S}(F_T)$ 
and ${\cal S}^*(F_T)$
characterized by their respective empirical cdf.   

The climatological forecaster is the only forecaster satisfying
\begin{equation}\label{eq: crps rv clim}
\mathrm{CRPS}(G,Y)  \stackrel{d}{ =  }  {\cal S^*}(G)  \stackrel{d}{ =  }  {\cal S}(G)\; .
\end{equation}
as by definition it discards any information about the system conditioning.
The first equality in (\ref{eq: crps rv clim}) is a direct consequence of auto-calibration, see \ref{framework-calib}; the second equality follows from the permutation invariance of the data from the point of view of the climatological forecaster.
 
\begin{figure*}
  \noindent\includegraphics[width=\textwidth,angle=0]{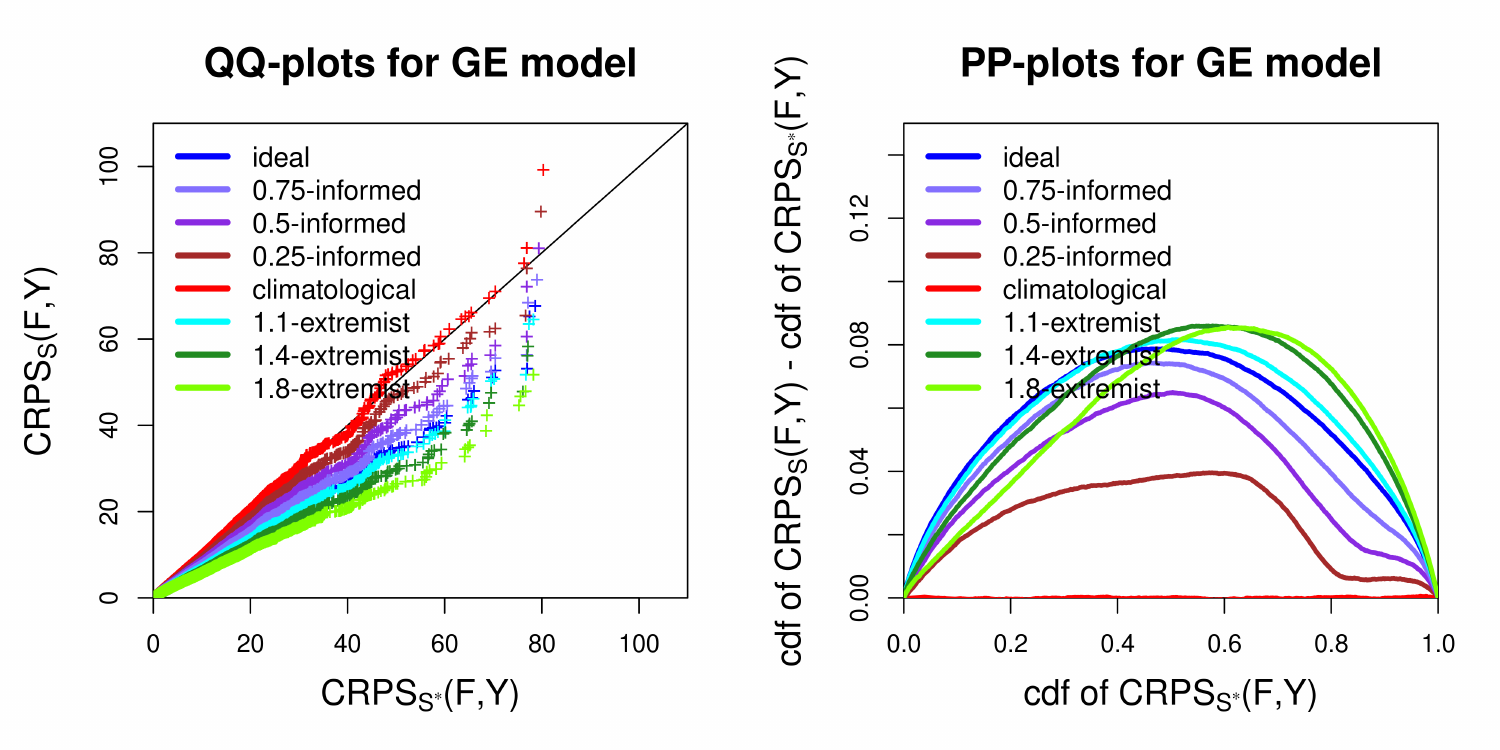}\\
  \caption{Comparisons of the distributional properties between ${\cal S}$ and  ${ \cal S^*}$ for each forecast in model GE:  
  qq-plots (left) and a pp-plots  (right panel). Each forecasts are represented by a sample of size $T=10^6$.
  }\label{qqplot}
\end{figure*}

\begin{table*}
\caption{Availability  status of the quantities of interest. It can be an a posteriori availability.
} 
\label{avail}
\begin{center}
~ \hskip -2cm
\begin{tabular}{ccc}

\hline
\hline
   Object & Definition & Availability \\
   & & in practice \\
 \hline
 $F_t$  & Distribution of the forecast for time $t$ & yes \\
 $y_t$ & Observed realisation at   time $t$ & yes \\
 $\delta_t$ & Conditioning variable & no \\
 $\Delta$ & Conditioning random variable & no \\
 $Y_t$ & Conditional random variable generating $y_t$ & no \\
 $Y$ & Unconditional random variable of the observations & yes \\
 $\mathrm{CRPS}(F_t,y_t)$ & CRPS of the couple for time $t$ & yes \\
 $\mathrm{CRPS}(F_t,Y_t)$ & Random variable associated to $CRPS(F_t,y_t)$ & no \\
 $\mathrm{CRPS}_{\cal S}(F,Y)$ & Random variable generated by the $(CRPS(F_t,y_t))_t$ & yes \\
  $\mathrm{CRPS}_{\cal S^*}(F,Y)$ & Random variable generated by the $(CRPS(F_t,y_{\pi(t)}))_t$  & yes \\
\hline
\end{tabular}
\end{center}
\end{table*}
The distributional properties of ${\cal S}(F_T)$, ${\cal S^*}(F_T)$, and ${\cal S}(G)$
give relevant insights on the behavior of the forecaster.
For illustration, Figure~\ref{qqplot} gives qq-plots of
the distributions of ${\cal S^*}(F_T)$ against ${\cal S}(F_T)$ for each forecast of the benchmark with $\gamma = 1/4$.  
We observe that the ideal, $\lambda$-informed and extremist forecasts deviate from the diagonal, illustrating the influence of the loss of information caused by the permutation:
such a visual diagnostic summarizes how ${\cal S}(F_T)$ and ${\cal S^*}(F_T)$ capture relevant information from the conditioning modelled here by the random variable $\Delta$.
The right panel of Figure~\ref{qqplot} displays these distributions on the probability scale and highlights how the discrepancy of the $\lambda$-informed forecaster evolves with the parameter $\lambda$.
Extremist forecasts, with multiple values of the scale parameter $\nu$, are displayed here for the sole purpose to illustrate how such visual diagnostics behave when calibration is not satisfied.
In Figure~\ref{qqplot}, we can also see that forecast dominance among forecasters could be inferred, as in \citet[][Fig. 1,2,4,6]{ehm2016quantiles} for point forecasts. Under calibration, discrepancy between distributions can be appropriately interpreted as a direct measure of the forecaster skill (the $\lambda$-informed curves never cross each other), making such diagnosis particularly relevant and compliant with the recommendations on the extremal dependence indices established by \citet{ferro2011extremal}.


\subsection{Tail properties of the random CRPS}
We now study  the upper tail behavior of the random CRPS, using EVT to develop a meaningful forecast evaluation for extreme events.
To lighten the technicality of this section, all proofs are relegated to \ref{app6p6}. 
In terms of notations with respect to any conditional model that depends on $\Delta = \delta$, we want to emphasize the difference between a conditional forecast, say $F_{\delta}$, and an unconditional forecast $F$. Note that $\delta$ depends on  the time index $t$, but for notation simplicity, we drop this index; $\Delta$ might also change over time but here assumed invariant.  

Let $X$ and $Y$ be two random variables with absolutely continuous cdfs $F$ and $G$ with common upper bound $x_{F}=x_{G}$.
Suppose that there exists 
$\gamma <1$ such that $G \in \mathcal{D}(H_{\gamma})$ and that $c_{F}=2\mathbb{E}_{F}(XF(X))$ is finite. Then conditionally on $\Delta=\delta$, one has
\begin{equation}\label{eq: conv}
\mathbb{P}\left.\left(\frac{\mathrm{CRPS}(F_\delta,Y_\delta)+c_{F_\delta}-u_\delta}{b_\delta(u_\delta)}>x \;\right|\;  Y_\delta>u_\delta\right)
\longrightarrow (1+\gamma_\delta x)^{-1/\gamma_\delta} \; ,
\end{equation}
as $u_\delta$ tends to $x_{G_\delta}$, with $1+\gamma_\delta x >0$.
So at any fixed state $\delta$ (state of the atmosphere for a weather forecast, say), the CRPS upper tail behavior (conditionally on $\Delta=\delta$) is equivalent to the observation tail behavior and formalizes what could be intuited from  (\ref{eq:wCRPS}). 

Now, unconditionally, one  can also get a result for the climatological forecast, thanks to its property of invariance under permutation (see Section \ref{sec:calibinfo}).
If there exists $\gamma <1$
such that $G\in \mathcal{D}(H_{\gamma})$, then 
 \begin{equation}\label{eq:unco}
  \mathbb{P}\left.\left\{\frac{\mathrm{CRPS}(G,Y)+c_{G}
  -u}{b(u)}> x \right| Y>u \right\}\longrightarrow (1+\gamma x)^{-1/\gamma}, \quad u\rightarrow x_{G},
 \end{equation}
for any $x$ such that $1+\gamma x >0$.
In the case where $\gamma >0$, convergence in Equation (\ref{eq:unco}) also holds for $c_{G} =0$ as the latter vanishes due to the linear behavior of the auxiliary function $b$ in Equation (\ref{eq: domain attraction}), e.g., see \citet[][]{embrechts1997modelling}. 


The benchmark presented in Table \ref{GEforecasters}  illustrates these results.
The choice of working  with a time indexed couple $(F_t, Y_t)$ or with an invariant $(G, Y)$ impacts significantly the tail behavior of the CRPS random variables:
according to Table \ref{GEforecasters}, the former case implies that  the  limit in (\ref{eq: conv}) exhibits an exponential tail, whereas the climatological tail given by (\ref{eq:unco}) is heavy, i.e., $\gamma > 0$.  

 \subsection{Assessing the forecaster tail behavior}\label{sec:index}
In this section, we propose a tail-equivalent forecast performance index inspired from
equations~(\ref{eq: conv}),  ~(\ref{eq:unco}), and Figure~\ref{qqplot}.
We aim only to provide the intuition behind the index and leave formal theoretical analysis for future work.
We assume that the forecasts lie in the domain of attraction of some distribution $H_{\gamma,\sigma}$. For sufficiently large $u$,  
the null hypothesis $H_{0} :  \mathcal{S}(F_T) | Y > u \stackrel{d}{ =  }  H_{\gamma, \sigma_u}$ should be rejected for any calibrated forecast with tail behaviour closer to the ideal forecast than the climatological reference.
To go further,  assume that the variables in  $\mathcal{S}(F_T)$ are iid. This assumption may not be always satisfied, as for instance temperature measures of two consecutive days are likely to be dependent, but can be reasonably satisfied for measurements from sufficiently far apart.
For each forecast, we can compute 
 a Cram\'er-von Mises criterion 
$$
{{\omega_u}}^2\{\mathcal{S}(F_T)\}=\int_{-\infty}^{+\infty} [{\hat{K}^{(m)}}_{\mathcal{S},u}(v)-H_{\gamma,\sigma_u}(v)]^2dH_{\gamma,\sigma_u}(v),
$$
where ${\hat{K}^{(m)}}_{\mathcal{S},u}$ is the empirical distribution of  the observations  in $\mathcal{S}(F_T)$ exceeding the threshold $u$.
The empirical nature of ${\hat{K}^{(m)}}_{\mathcal{S},u}$ allows to simplify ${\omega_u}^2\{\mathcal{S}(F_T)\}$ to
$$
{\Omega^F_u} = m \times \widehat{\omega_u}^2\{\mathcal{S}(F_T)\} = \frac{1}{12m} + \sum_{i=1}^m \left[\frac{2i-1}{2m}-H_{\gamma,\sigma_u}(s_i)\right]^2,
$$
where $m$ denotes the number of observations exceeding $u$ and $s_1,\dots,s_m$ are the ordered values of $\mathcal{S}(F_T)$.
{A detailed algorithm for the computation of $\Omega^F_u$ is provided in Table \ref{algo} of \ref{app: algo}.}


%
%
As suggested by Figure~\ref{qqplot}, we assume that $\Omega^F_u > \Omega^G_u$, for any calibrated forecasts and  climatology $G$. Also, for two calibrated forecasts $F^1$ and $F^2$, we conjecture that $\Omega^{F^2}_u \geq \Omega^{F^1}_u $ if $F^2$ has a tail behaviour closer to the ideal forecast than $F^1$. Under these assumptions, we can summarize simply the comparison between $\Omega^F_u$ and $\Omega^G_u$ through
\begin{equation}\label{eq: index}
T_u(F,G)= 1- \frac{\Omega^G_u}{\Omega^F_u} .
\end{equation}

\begin{figure*}[!h]
  \noindent\includegraphics[width=\textwidth,angle=0]{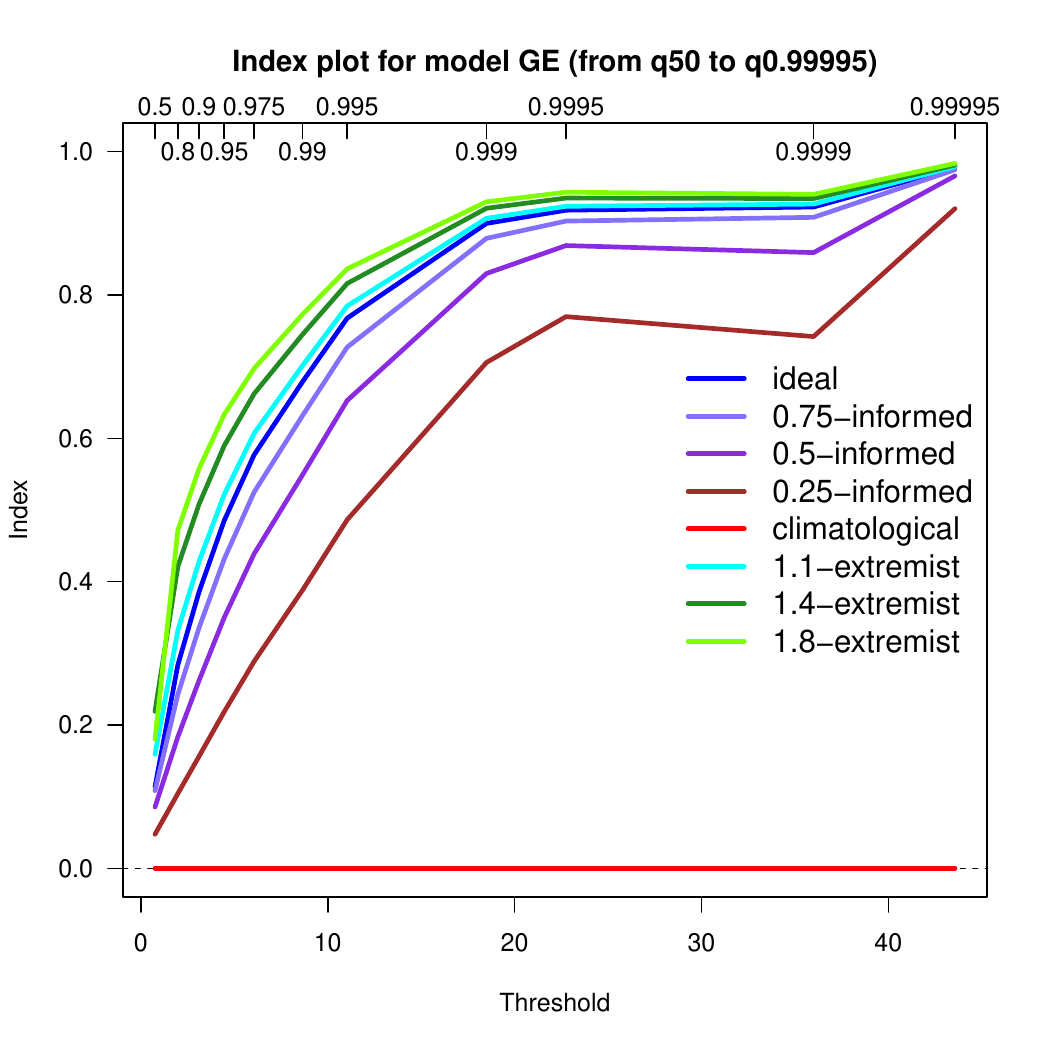}\\
  \caption{Cram\'er-von Mises' criterion-based index as a function of the threshold for the different forecasts in model GE with parameters $T = 10^6$ and $\gamma = 1/4$. Indexes are computed for thresholds ranging from the 0.5 to the 0.99995 empirical quantile. Higher index values are assumed to reflect a tail behaviour closer to the ideal forecaster. Validity of the index is limited to calibrated forecast and Non-calibrated extremists forecast are shown to recall that calibration must be first carefully checked before interpreting such graphics.}\label{pva}
\end{figure*}

The behaviour of the index $T_u$ is illustrated with the help of model GE;
Figure \ref{pva} displays the evolution of $T_u$ as a function of the threshold $u$ for $T=10^6$ and $\gamma = 1/4$.
The behaviour of the index is shown to be consistent with our conjecture:  first, the ideal forecast performs best, while the climatology has the lowest index.
Performance ranking among calibrated forecasters is stable as the threshold increases, with the ideal forecast always obtaining the largest index.
The extremist forecasters, displayed here to illustrate the behaviour of the index for non-calibrated forecast, obtain a high index, even larger than the ideal forecast, stressing the importance of calibration which must be carefully assessed before any interpretation of $T_u$.

In practice, a threshold choice has to be made, for which numerous methodologies have been developed, see, e.g., \citet{beirlant:2004, papastathopoulos2013extended, wrcrneaveau}.


\section{Discussion}\label{blah}

In this work, we have argued with the help of a carefully designed benchmark that the mean of the CRPS, or its weighted counterparts, are unable to successfully discriminate a forecast upper tail regime, as demonstrated by \citet{brehmerStrokorb19}.
\citet{ehm2016quantiles} have introduced the so-called ``Murphy diagrams'' for assessing dominance in point forecasts.
This original approach allows to appreciate dominance among different forecasts and anticipate their skill area; a similar visual diagnostic is presented in Figure \ref{qqplot} for calibrated forecasts.

Inspired by \citet{friederichs2012forecast}, we apply EVT directly on common verification measures.
By considering the CRPS  as a random variable, see also \cite{Bessac21} for non-extreme cases, 
one can view this contribution as a first step in considering other functionals of the scores distributions rather than their means.
The new index introduced in Section~\ref{sec:index} can be considered as a probabilistic alternative to the scores introduced by \citet{ferro2007probability} and \cite{ferro2011extremal}. 
We make a link between the paradigm of \textit{maximizing the sharpness subject to calibration} from \citet{gneiting2007probabilistic} and the paradigm of \textit{maximizing the information for extreme events subject to calibration}.
In a same vein, \citet{murphy1993good} has presented the differences between forecast quality (accordance between forecasts and observations) and forecast value (ability to bring information to realize a benefit by choosing a forecast),  the forecast value seems to be the most important for extreme events, where decision making is crucial. For deterministic weather forecasts, such tools are well-known, see e.g. \citet{richardson2000skill,zhu2002economic}. 
Other widely-used scores based on the dependence between forecasts and observed events have been considered in \citet{stephenson2008extreme,ferro2011extremal}.

It would be worthwhile to further study the theoretical properties of this CRPS-based tool.
Another potentially interesting investigation could be to extend this procedure to other scores like the mean absolute difference, the Dawid-Sebastiani score \citep{dawid1999coherent} or the ignorance score \citep{smith2015towards,diks2011likelihood}. Classical tools in verification relies on a verification period, as a consequence evaluation is always done a posteriori. Thus, an interesting manner to pursue this work would be to consider sequential evaluation of rare events, in the spirit of the e-values \citep{vovk2021values} introduced to assess and monitor calibration continuously \citep{arnold2021sequentially}.
Eventually, we invite scientists to work on new theory of scoring rule departing from the score's averages.

\section*{Acknowledgments}

Part of this work was supported by the French National Research Agency (ANR) project
T-REX (ANR-20-CE40-0025) and by Energy oriented Centre of Excellence-II (EoCoE-II), Grant Agreement 824158, funded within the Horizon2020 framework of the European Union. Part of this work was also supported by the ExtremesLearning grant from 80 PRIME CNRS-INSU and the ANR project Melody (ANR-19-CE46-0011). This work was partially supported by the ANR LABEX MILYON (ANR-10-LABX-0070) of Université de Lyon, within the program "Investissements d'Avenir" (ANR-11-IDEX-0007). 

\section*{Implementation details}
The implementation of the index relies on the \texttt{extremeIndex} package \citep{rpack}. The \texttt{R} code generating simulation data and Figures is available upon request.

\bibliographystyle{elsarticle-harv} 
 \bibliography{references}


\appendix
\section{Prediction framework and calibration} \label{framework-calib}
%
%
%
 The theoretical framework considered in this paper is the now classical \textit{prediction space} already introduced by \citet{murphy1987general,gneiting2013combining,ehm2016quantiles}, and generalized in a serial context by \citet{strahl2017cross}. It starts formally with a probability space $(\Omega, {\mathcal A}, {\mathbb Q})$ and a  collection of sub-$\sigma$-algebras $\mathcal{A}_1, \dots, \mathcal{A}_k \subset {\mathcal A}$, where $\mathcal{A}_i$  represents the information available to forecaster $i$. 
 In a meteorological context, it can be seen as the representation of the atmosphere done by each forecaster. In the benchmark considered in Section~\ref{sec:GEmodel}, we will consider for simplicity that the information set 
 is generated by a random variable $\Delta$.
 
 A real-valued outcome $Y$ is observed and seen as a (real-valued) random variable. 
  A probabilistic  forecast $i$ for  $Y$ is identified with its so-called ``predictive distribution'' with cdf $F_i$. Rigorously speaking, $F_i: \Omega \times {\mathcal B}({\mathbb R}) \to [0,1]$ is a kernel\footnote{This means that for each fixed $\omega \in \Omega$, $F_i(\omega, \cdot)$ is a probability measure, and for each fixed $x \in {\mathbb R}$, $F_i(\cdot, (-\infty, x])$ is 
  $ {\mathcal A}_i$-measurable. See e.g. Kallenberg (2017).} from    $(\Omega, {\mathcal A}_i)$ to $({\mathbb R}, {\mathcal B}({\mathbb R}))$, but as done by previous authors, we will identify  the kernels with random cumulative cdf, see e.g.  \citet{strahl2017cross} for more details. For each $x \in {\mathbb R}$, we might in particular use the notation 
  $F_i(x)$ meaning the random element $\omega \mapsto F_i(\omega,  (-\infty, x])$.
  
In such a framework, a forecast $F_i$ is termed {\it ideal} with respect to ${\mathcal A}_i$ if $F_i = {\mathcal L}(Y | {\mathcal A}_i)$ almost surely. 
\citet{tsyplakov2011evaluating} also refers to this property saying that $F_i$ is  {\it calibrated} with respect to ${\mathcal A}_i$. He additionally 
defines the {\it auto-calibration} as the property for $F_i$ to satisfy  $F_i = {\mathcal L}(Y | \sigma(F_i))$ almost surely. Here, $\sigma(F_i)$ denotes the $\sigma$-algebra generated by $F_i $, that is to say the smallest $\sigma$-algebra such that $\omega \mapsto F_i(\omega, x)$ is measurable for all $x \in {\mathbb R}$.
Note that if a forecast is calibrated with respect to ${\mathcal A}_i$, then it is auto-calibrated, but the converse does not hold  in general.
As a particular case considered in Section~\ref{sec:GEmodel},  the \textit{climatological} forecaster 
is ideal with respect to the trivial $\sigma$-algebra.
  
%
%
%

%
   
 In practice, one is not only concerned with predictions for an outcome $Y$ at a single time point. The framework introduced above also allows to deal with independent replicates at times $t=1, 2, \dots$, as is done in Section~\ref{sec:GEmodel}.  If such an assumption of independence sounds  unrealistic in several situations, as argued by  \citet{strahl2017cross}, it can nevertheless provide a first step and   takes advantage of  a lighter context. We chose therefore to keep it in this paper for simplicity.

\section{An alternative  expression of the weighted CRPS} \label{app:CRPSequivalences}
The weighted CRPS defined by (\ref{eq:def-wCRPS}) can be reformulated in the following way, as soon as the weight function $w(.)$ is continuous,
\begin{equation} \label{eq:wCRPS}
wCRPS(F,y)  = W(y)+2\mathbb{E}_F[\{W(X)-W(y)\}\mathbf{1}_{X > y}] 
-2\mathbb{E}_F[W(X)F(X)] \; .
\end{equation} 
Assume that the weight function $w(.)$ is continuous. 
By integrating by parts $\int_{-\infty}^{y} F^2(x) w(x)\,dx$ and $\int_{y}^{\infty} \overline{F}^2(x)  w(x)\,dx$ and 
using $W(x) =\int_{-\infty}^x w(z) dz$, the weighted CRPS defined by    (\ref{eq:def-wCRPS})
can be rewritten as 
$$
wCRPS(F,y)=\mathbb{E}_F|W(X)-W(y)|-\frac{1}{2}\mathbb{E}_F|W(X)-W(X')|.
$$ 
The equality $|a-b| = 2 \max(a,b)- (a+b)$ gives 
\begin{eqnarray*} 
 \mathbb{E}_F|W(X)-W(y)| 
 &=&2 \mathbb{E}_F \max(W(X),W(y))- \mathbb{E}_F W(X) - W(y),\\
 &=&W(y)  -  \mathbb{E}_F W(X) + 2 \mathbb{E}_F \left(W(X)-W(y)  I[W(X) >  W(y)] \right),
\end{eqnarray*}
and
\begin{eqnarray*} 
 \mathbb{E}_F|W(X)-W(X')| &=&2 \mathbb{E}_F \max(W(X),W(X'))- 2 \mathbb{E}_F W(X),\\
 &=&4  \mathbb{E}(W(X)F_{W(X)}(W(X))) - 2 \mathbb{E}_F W(X),\\
 &=&4  \mathbb{E}(W(X)F(X)) - 2 \mathbb{E}_F W(X) \; ,
\end{eqnarray*}
where the last line follows from the fact that  $F_{W(X)}(W(X))$ and  $F(X)$ have the same distribution, which is uniform on $(0,1$).
As $W(x)$ is non-decreasing, one has  $\{ W(X) >  W(y)\} = \{ X >  y \}$, and it follows that 
\begin{eqnarray*} 
\mathrm{wCRPS}(F,y) &= & W(y)  -  \mathbb{E}_F W(X) + 2 \mathbb{E}_F \left[\{W(X)-W(y)\}  \mathbf{1}_{W(X) >  W(y)} \right] \\
& & - 2 \mathbb{E}_F[W(X)F(X)] + \mathbb{E}_F W(X), \\
 &=& W(y)+2\mathbb{E}_F[\{W(X)-W(y)\}\mathbf{1}_{X > y}] 
-2\mathbb{E}_F[W(X)F(X)] \; ,
\end{eqnarray*} 
as announced in (\ref{eq:wCRPS}).
%

\section{Proof of the inequality  (\ref{eq: tail equiv})} \label{app:ineq-wCRPS}



 Let $u$ be a positive real. Denote $Z$ a non-negative random variable with finite mean and cdf $H$. Assume that $Z$ and $Y$ are independent and have same right end point.
  We introduce the new random variable
 \begin{equation}
  X_u =Y\mathbf{1}\{u\geq Y\}+(Z+u)\mathbf{1}\{Y > u\} \; ,
 \end{equation}
with survival function $\overline{F_u}$ defined by 

\begin{eqnarray}\label{cdftail}
 \overline{F_u}(x) &=& \left\{ \begin{array}{ll}
   \overline{G}(x), & \; \;  \mbox{if } x\leq u \\
  \overline{H}(x-u)\overline{G}(u), & \; \; \mbox{otherwise}.
  \end{array}                 
             \right.
\end{eqnarray}
Note that the decreasingness of $\overline{F_u}$  yields in particular that for all $x$,
\begin{equation}
 \overline{F_u}(x)\leq \overline{G}(x) \; .
\end{equation}
Besides, equation (\ref{cdftail}) and the monotonicity of
$W$ allows to write that for any $x\leq u$

\begin{equation}\label{eqesp}
 \mathbb{E}[W(Y)\mathbf{1}\{Y<x\}]=\mathbb{E}[W(X_u)\mathbf{1}\{X_u<x\}]\; .
\end{equation}
Equality (\ref{eq:wCRPS}) 
implies that
\begin{eqnarray*}
&& \hskip -1.2cm \frac{1}{2}[\mathrm{wCRPS}(F_u,x)  -\mathrm{wCRPS}(G,x)]  \\
 &=&\mathbb{E}_{F_u}[(W(X_u)-W(x))\mathbf{1}\{X_u>x\}] -\mathbb{E}_G[(W(Y)-W(x))\mathbf{1}\{Y>x\}]\\
 &&+\mathbb{E}_G[W(Y)G(Y)]-\mathbb{E}_{F_u}[W(X_u)F_u(X_u)],\\
 &=&\mathbb{E}_{F_u}[W(X_u)\overline{F_u}(X_u)]-\mathbb{E}_G[W(Y)\overline{G}(Y)]\\
 &&-\mathbb{E}_{F_u}[(W(X_u)-W(x))\mathbf{1}\{X_u\leq x\}]+\mathbb{E}_G[(W(Y)-W(x))\mathbf{1}\{Y\leq x\}]\\
 &=&\mathbb{E}_{F_u}[W(X_u)\overline{F_u}(X_u)]-\mathbb{E}_G[W(Y)\overline{G}(Y)]+ \Delta(x) \; , 
\end{eqnarray*}
where
$$
 \Delta(x)=\mathbb{E}_G[(W(Y)-W(x))\mathbf{1}\{Y\leq x\}]-\mathbb{E}_{F_u}[(W(X_u)-W(x))\mathbf{1}\{X_u\leq x\}].
$$
The stochastic ordering that holds between $X_u$ and $Y$ implies that the quantity $\mathbb{E}_{F_u}[W(X_u)\overline{F_u}(X_u)]-\mathbb{E}_G[W(Y)\overline{G}(Y)]$ is negative. 
 Combined with (\ref{eqesp}), this leads to
\begin{eqnarray}\label{eq:auxDelta}
 \frac{1}{2}\left|\mathbb{E}_G[\mathrm{wCRPS}(F_u,Y)]-\mathbb{E}_G[\mathrm{wCRPS}(G,Y)]\right|\leq \int_u^{x_G}\Delta(x) dG(x).
\end{eqnarray}
For $x>u$ we can write that
\begin{eqnarray*}
 && \hskip -1cm \Delta(x)\\
 &=&\mathbb{E}_Y[(W(Y)-W(x))\mathbf{1}\{u<Y\leq x\}]-\mathbb{E}_{F_u}[(W(X_u)-W(x))\mathbf{1}\{u<X_u\leq x\}],\\
 &\leq& \mathbb{E}_{F_u}[(W(x)-W(u))\mathbf{1}\{u<X_u\leq x\}] ,
 \end{eqnarray*}
  since $W(Y)-W(x)\leq 0$ in the first expectation, whereas $0\leq W(x)-W(X_u) \leq W(x)-W(u)$ in the second one. As a consequence, one gets
 \begin{eqnarray*}
 \Delta(x) & \leq &  (W(x)-W(u))[F_u(x)-F_u(u)],\\
 &\leq& (W(x)-W(u))\overline{F_u}(u),\\
 &=&(W(x)-W(u))\overline{G}(u).
\end{eqnarray*}
This last expression combined with (\ref{eq:auxDelta}) leads finally to
\begin{eqnarray*}
\hskip -5mm  \left|\mathbb{E}_G[\mathrm{wCRPS}(F_u,Y)]-\mathbb{E}_G[\mathrm{wCRPS}(G,Y)]\right| &\leq& 2\overline{G}(u)\int_u^{x_G}(W(x)-W(u)) dG(x) .
\end{eqnarray*}
Note that this inequality is true for any $u$ and $H$, and its right hand side does not depend on $\overline{H}(x)$. Thus, 
the tail behavior of the random variables $Y$ and $Z$ can be completely different, although the CRPS of $G$ and $G$ can be as closed as one wishes. 
The right hand side goes to $0$ due to the finite mean of $W(Y)$.

\section{A detailed example related to Section 2.2} \label{app:ex-2.2}
In this appendix, we illustrate the fact that the CRPS fails at discriminating forecasts with different tails. 
We consider GP distributed  forecasts and observations.
In this case, closed form of the CRPS are available, as detailed in the following.  
%
\begin{lemma}\label{lemma:crps}
Consider $X \stackrel{d}{ =  } \mbox{\rm GP}(\beta, \xi)$ and $Y \stackrel{d}{ =  } \mbox{\rm GP}(\sigma, \gamma)$ with $ 0 \leq \xi <1$ and $0 \leq \gamma <1$, 
with respective survival functions $\overline{F}(x) = (1 + \xi x/\beta)^{-1/\xi}$ (for $x > -\beta/\xi$) and $\overline{G}(x) = (1 + \gamma x/\sigma)^{-1/\gamma}$ (for $x > -\sigma/\gamma$). 
If  $\gamma/\sigma= \xi/\beta$, 
with $\gamma \neq 0$, then 
$$ 
\mathbb{E}_{G}\left[ \mathrm{CRPS}(F,Y) \right]=\frac{\sigma}{1- \gamma}  +2 \beta  \left[  \frac{1}{2(2-\xi)}- \frac{\gamma}{\gamma + \xi-\gamma\xi}\right].
$$
This gives the minimum \textrm{CRPS} value for $\xi=\gamma$ and $\sigma=\beta$,
$$
\mathbb{E}_{G}\left[ \mathrm{CRPS}(G,Y) \right] = \frac{\sigma}{(2-\gamma)(1-\gamma)}.
$$
\end{lemma} 
{\bf Proof:}
Applying (\ref{eq:wCRPS}) with $W(y)=y$, and making use of classical properties of the Pareto distribution (see e.g. \cite[Theorem 3.4.13]{embrechts1997modelling}), one gets 
\begin{equation}\label{eq: crps(F,y) GP}
\mathrm{CRPS}(F,y) = y +2(1 + \xi y/\beta)^{-1/\xi} \frac{\beta + \xi y}{1-\xi}-2 \beta \left( \frac{1}{1-\xi}  -\frac{1}{2(2-\xi)}  \right).
\end{equation}
It follows that 
$$
\mathbb{E}\left[ \mathrm{CRPS}(F,Y) \right] =  \frac{\sigma}{1- \gamma}  +2 \frac{\beta }{1-\xi} m_0
+ 2 \frac{\xi }{1-\xi} m_1
  -2 \beta \left( \frac{1}{1-\xi}  -\frac{1}{2(2-\xi)}  \right),
$$
with  
$$
m_0 =  \mathbb{E}\left[  \left(1 + \frac{\xi}{\beta} Y\right)^{-1/\xi}  \right], \mbox{ and } 
m_1 =  \mathbb{E}\left[  Y \left(1 + \frac{\xi}{\beta} Y\right)^{-1/\xi}   \right].
$$ 
%
Since
$$
 \left(1 + \frac{\xi}{\beta} y\right)^{-1/\xi} 
 = \overline{G}^s \left(  c y\right), \mbox{ with } c=\frac{\xi \sigma}{\beta \gamma} \mbox{ and } s=\frac{\gamma}{\xi},  
$$
one can write 
$$
m_r =  \mathbb{E}\left[  Y^r \overline{G}^s \left(  c Y\right)  \right]  \mbox{ for } r=0,1.
$$ 
Besides, as $G^{-1}(v)= \frac{\sigma}{\gamma} \left(  \left( 1-v\right)^{-\gamma} -1\right)$, 
one can thus rewrite,  denoting by  $U$ a random variable uniformly distributed on $(0,1)$,  
\begin{eqnarray*} 
m_r &=&   \mathbb{E}\left[  G^{-1}(U)^r \overline{G}^s \left(  c G^{-1}(U)\right)  \right],\\
&=& \mathbb{E}\left[  \left( \frac{\sigma}{\gamma} \left(  \left( 1-U\right)^{-\gamma} -1\right) \right)^r \left(1+ \frac{\gamma}{\sigma} \left(  c \frac{\sigma}{\gamma} \left(  \left( 1-U\right)^{-\gamma} -1\right) \right) \right)^{-s/\gamma}  \right],\\
 &=&  \left( \frac{\sigma}{\gamma}\right)^r   \mathbb{E}\left[   \left(  U^{-\gamma} -1\right)^r \left((1-c) + c U^{-\gamma}\right)^{-s/\gamma}  \right],\\
 &=&  \left( \frac{\sigma}{\gamma}\right)^r   \mathbb{E}\left[   \left(  \frac{B}{1-B}\right)^r \left(\frac{1-(1-c)B}{1-B}\right)^{-s/\gamma}  \right], \mbox{ with } B= 1-U^{\gamma}\\
 &=&  \left( \frac{\sigma}{\gamma}\right)^r   \mathbb{E}\left[  B^r (1-B)^{-r + s/\gamma}  \left(1-(1-c)B\right)^{-s/\gamma}  \right], \mbox{ with } B\sim \mbox{Beta}(1, 1/\gamma)\\
 &=&  \left( \frac{\sigma}{\gamma}\right)^r   \mathbb{E}\left[  B^r (1-B)^{-r + 1/\xi}  \left(1-(1-c)B\right)^{-1/\xi}  \right], \mbox{ because } s/\gamma=1/\xi.\\
\end{eqnarray*} 
If $\displaystyle c= \frac{\xi \sigma}{\beta \gamma}=1$, then this simplifies to
\begin{eqnarray*} 
m_r &=& \left( \frac{\sigma}{\gamma}\right)^r   \frac{1}{\gamma}\int_0^1 u^r (1-u)^{-r +  1/\xi + 1/\gamma -1} du =   \left( \frac{\sigma}{\gamma}\right)^r \frac{1}{\gamma} B(r+1, -r + 1/\xi + 1/\gamma),\\
&=&   \left( \frac{\sigma}{\gamma}\right)^r \frac{1}{\gamma}  \frac{\Gamma(r+1)\Gamma(-r + 1/\xi + 1/\gamma)}{\Gamma(1+ 1/\xi + 1/\gamma)}.
\end{eqnarray*} 
In particular, 
$m_0=  \frac{1}{\gamma} B(1,  1/\xi + 1/\gamma) =  \left( 1+ \frac{\gamma}{\xi}\right)^{-1}$ and 
$$
m_1= \frac{\sigma}{\gamma}   \left( 1+ \frac{\gamma}{\xi}\right)^{-1}  \left( \frac{1}{\xi}+ \frac{1}{\gamma}-1\right)^{-1}. 
$$
It follows that, if $\frac{\gamma}{\sigma}= \frac{\xi}{\beta}$, then we have 
\begin{eqnarray*} 
\mathbb{E}\left[\mathrm{CRPS}(F,Y) \right] &=  &\frac{\sigma}{1- \gamma}  +2 \beta  \left[  \frac{1}{2(2-\xi)}- \frac{\gamma}{\gamma + \xi-\gamma\xi}\right].
\end{eqnarray*} 
This gives the minimum CRPS value for $\xi=\gamma$ and $\sigma=\beta$, 
$$
\mathbb{E}\left[ \mathrm{CRPS}(G,Y) \right] = \frac{\sigma}{(2-\gamma)(1-\gamma)},
$$
concluding the proof of Lemma~\ref{lemma:crps}.
\hfill $\square$

Lemma \ref{lemma:crps} allows to  study the effect of changing the forecast's  tail behavior captured by $\xi$ and the spread forecast encapsulated  in $\beta$, when $F$ and $G$ have proportional parameters, i.e., $\beta =a \sigma$ and $\xi =a \gamma$ for some $a>0$. 
In this case,  the CRPS simplifies to
\begin{equation}\label{eq: Pareto CRPS}
\mathbb{E}_{G}\left[ \mathrm{CRPS}(F,Y) \right] =\frac{\sigma}{1- \gamma}  +2 a \sigma  \left[  \frac{1}{2(2-a \gamma)}- \frac{1}{1+a - a \gamma}\right] \; ,
\end{equation}
leading when $a>1$ to a  forecaster with  heavier-tail,  overestimating the true upper tail behavior, and to the  opposite when $a<1$. 

Counter examples as the previous one can thus be found, illustrating 
how weighted scoring rules fail to compare tail behaviors. They should therefore be handled with a particular care, especially for forecast makers, as already advocated by \citet{gille18,lerch2017forecaster}.

\section{Proof of the convergences  (\ref{eq: conv}) and (\ref{eq:unco})}   \label{app6p6}
The proof of  (\ref{eq:unco}) can be seen as a particular case of (\ref{eq: conv}),  so that we will focus on proving (\ref{eq: conv}). The following lemma will help to get the result, and is presented first with its proof. In what follows, the mean excess  function of any random variable $Z$ with finite mean and with cdf $F$ will be denoted by  $M(F,z)$, so that  $\overline{F}(z) M(F,z)=\mathbb{E}_F[(Z-z){ \rm{1}\!l}_{Z>z}].$ 
\vskip 2mm \noindent
{\bf Lemma :}
 Consider a random variable $Z$ with finite mean that belongs to domain of attraction $\mathcal{D}(H_\gamma)$ with $\gamma<1$. 
 There exist non negative real numbers $\alpha$ and $\beta$ such that for each $z \in {\mathbb R}$,
 \begin{equation}\label{eq:lemma}
  0 \leq 2\mathbb{E}_F\left[(Z-z) { \rm{1}\!l}_{Z>z}\right]\leq  \overline{F}(z)(\alpha z +\beta) \; .
 \end{equation}

%

 
\vskip 2mm \noindent 
{\it Proof of the lemma:}
The indicator function ${ \rm{1}\!l}_{Z>z}$ implies that we always have $0 \leq 2\mathbb{E}_F((Z-z) { \rm{1}\!l}_{Z>z})$. 
To prove that $2\mathbb{E}_F((Z-z) { \rm{1}\!l}_{Z>z})$ is smaller than $  \overline{F}(z)(\alpha z +\beta)$, we first show that this inequality  holds for large values of $z$.  Note first that if $z > x_F$, then (\ref{eq:lemma}) is trivially true. Let then show the result when $z \stackrel{<~}{\rightarrow} x_F$,
and for this, let
  decompose the proof depending on the sign of $\gamma$ : 
 \begin{enumerate}
  \item  $F$ belongs to $\mathcal{D}(H_{\gamma})$ with $0<\gamma<1$ :
  In this case, \citet{embrechts1997modelling} (Section 3.4) show that  $M(F,z)\sim \gamma z/(1-\gamma)$ as $z$ tends to $x_F$, and we can conclude directly.
  \item  $F$ belongs to $\mathcal{D}(H_{\gamma})$ with $\gamma<0$ :
  In this case, the result also follows easily from \citet{embrechts1997modelling} since when   $ z$ tends to $x_F$, $M(F,z)\sim \gamma(x_F-z)/(\gamma-1).$ This allows to fix $\alpha =0$ and $\beta= \sup_{z \in V(x_F)} \gamma(x_F-z)/(\gamma-1)$ for an appropriate neighborhood $V(x_F)$ of $x_F$.
  
  \item $F$ belongs to  $\mathcal{D}(H_{0})$ : 
  When $F$ is in the Gumbel domain of attraction, $M(F,z)/z\rightarrow 0$ as $z$ tends to $x_F$ (see e.g. Theorem 3.9 in \citet{ghosh2010discussion}). 
  If  $x_F$ is finite, then there exists a positive $\beta$ such that $2M(F,z) \leq  \beta$ and $\alpha$ can be fixed to 0, whereas if  $x_F$ is infinite, the fact that $2M(F,z) < z$ for $z$ large enough enables to conclude.
 \end{enumerate}
So far, we have shown that,  for some large $z_0$, there exist non negative $\alpha$ and $\beta$ such that 
 $$
 2\mathbb{E}_F((Z-z) { \rm{1}\!l}_{Z>z})\leq  \overline{F}(z)(\alpha z +\beta) \mbox{, for all $z>z_0.$}
 $$
We still need to prove that this statement also holds for $z\leq z_0$. 
 Define 
 $$
 0 \leq  \beta_0 = 2 \max_{z\leq z_0} \mathbb{E}_F[(Z-z) { \rm{1}\!l}_{Z>z}]. 
 $$ 
 As $\gamma<1$, $\beta_0$ is finite and, as $\overline{F}(z) \geq  \overline{F}(z_0)$ for all $z\leq z_0$, we have 
 $$
 0 \leq \beta_0 \leq \beta_0 \frac{\overline{F}(z)}{\overline{F}(z_0)}. 
 $$
 We have now two cases:  either $\beta< \frac{\beta_0}{\overline{F}(z_0)}$ or $\beta \geq  \frac{\beta_0}{\overline{F}(z_0)}$. 
 In the latter case, we have $ 2\mathbb{E}_F((Z-z) { \rm{1}\!l}_{Z>z}) \leq \beta_0 \leq \overline{F}(z)(\alpha z +\beta)$, 
 and so, the required  result is obtained. 
 In the case of $\beta< \frac{\beta_0}{\overline{F}(z_0)}$, it is always possible to increase  $\beta$ chosen when $z>z_0$,  and bring it above $\frac{\beta_0}{\overline{F}(z_0)}$. \qed

\vskip 2mm
We are now ready to prove  (\ref{eq: conv}) as announced.
\vskip 2mm
\noindent
{\it Proof of (\ref{eq: conv}):}\\
Given the conditional  forecast $F_{\delta}$, the CRPS  can be computed with respect to the conditional observation $y_{\delta}$ in the following way 
$$
\mathrm{CRPS}(F_{\delta},y_{\delta})\stackrel{}{=} y_{\delta}-c_{\delta} + 2\mathbb{E}_{F_{\delta}}\left[(X_{\delta}-y_{\delta}) 1(X_{\delta}>y_{\delta})\right],
$$
where    $c_{\delta}=2\mathbb{E}_{F_{\delta}}\left[X_{\delta}F_{\delta}(X_{\delta})\right]$.
To simplify notations, we drop the subscript $\delta$ in the rest of the proof, but it  will be back at the end. The previous lemma allows to write 
$$
Y\leq \mathrm{CRPS}(F,Y)+c \leq (1+\alpha\overline{F}(Y))Y+\beta\overline{F}(Y) \;\;\; a.s.
$$
Let now work conditionally on  $Y > u$, for a large $u$ close to $x_F=x_Y$.
We then get 
$$
Y\leq \mathrm{CRPS}(F,Y)+c \leq (1+\alpha\overline{F}(u))Y+\beta\overline{F}(u) \;\;\; a.s.
$$
This holds when the right end point of $Y$ is non-negative. If this was not the case, note that one can  simply  write  
$Y\leq \mathrm{CRPS}(F,Y)+c \leq Y+\beta\overline{F}(u) \;\;\; a.s.$. 

The main idea of the proof is to notice that $\overline{F}(u)$ goes to zero as $u$ gets large, and consequently, the above inequalities indicate that 
the thresholded random variable 
$Y[u]=[(Y -u)/b(u)\;|\; Y>u]$  and the thresholded CRPS $C[u]= [(\mathrm{CRPS}(F,Y)+c -u)/b(u)\;|\; Y>u]$  should behave similarly  for 
large~$u$.
The choice of positive constant $b(u)$ depends on the domain of attraction of $Y$. 
More precisely, we assume that $Y[u]$ converges in distribution towards a GPD with finite mean. 
 So that
 \begin{eqnarray*}
&&  0\leq \mathbb{P}\left(\frac{\mathrm{CRPS}(F,Y)+c-u}{b(u)}>t \;|\;  Y>u\right)-\mathbb{P}\left(\frac{Y-u}{b(u)}>t \;|\;  Y>u\right) \\
&\leq&  \mathbb{P}([1+\alpha\overline{F}(Y)]Y+\beta\overline{F}(Y)>tb(u)+u \;|\; | Y>u)-\mathbb{P}(Y>tb(u)+u \;|\; Y>u)\\
&\leq&  \mathbb{P}\left(Y>\frac{tb(u)+u-\beta\overline{F}(u)}{1+\alpha\overline{F}(u)} \;|\;  Y>u\right)-\mathbb{P}(Y>tb(u)+u \;|\; Y>u).\\
  \end{eqnarray*}
 We recognize the probability (conditionally on $Y>u$) for $Y$ to be in an interval denoted by 
 $$
 I_u = \left[\frac{tb(u)+u-\beta\overline{F}(u)}{1+\alpha\overline{F}(u)},tb(u)+u\right].
 $$
 The remaining part of the proof consists in showing that this conditional probability tends to 0 as $u \to x_F$.
 We can write 
 $$
  \mathbb{P}\left(Y \in I_u \;|\; Y >u  \right) =  \mathbb{P}\left(Y \in u + J_u \;|\;  Y >u  \right)\; ,
 $$
 where $J_u = \displaystyle \left[\frac{tb(u)-\overline{F}(u)(\alpha + \beta)}{1+\alpha\overline{F}(u)},tb(u)\right].$
 For $u$ large enough, the latter probability can be approximated by a GPD, so that 
 \begin{eqnarray*}
  \mathbb{P}\left(Y \in I_u \;|\;  Y >u  \right)& \sim & |J_u|  \sup_{v \in J_u}g_{GP}(v)
  =  \frac{\overline{F}(u)[\alpha + \beta + \alpha t  b(u)]}{1+\alpha\overline{F}(u)} \sup_{v \in J_u}g_{GP}(v) \; ,
 \end{eqnarray*}
 where $g_{GP}$ denotes the probability density function associated to the GPD.
This implies the convergence to 0 of the latter probability. Since this is true conditionally on $\Delta = \delta$, it can be rewritten, after reintroduction of the subscript $\delta$, as
$$
\mathbb{P}\left(\frac{\mathrm{CRPS}(F_\delta,Y_\delta)+c_\delta-u_\delta}{b_\delta(u_\delta)}>x \;|\;  Y_\delta>u_\delta\right)
\longrightarrow (1+\gamma_\delta x)^{-1/\gamma_\delta} \; ,
$$
as $u$ tends to $x_{G_\delta}$, with $1+\gamma_\delta x >0$. \qed

\newpage 
\section{Algorithm for the computation of the Cramer-von-Mises criterion}\label{app: algo}

\begin{table*}[!h]
\caption{Computation of  Cram\'er-von Mises' statistic  from    $N$ couples forecast/observation. It can be done with the \texttt{R} package \texttt{extremeIndex} \citep{rpack}.}\label{algo}
\begin{center}
\begin{tabular}{p{0.35\textwidth}p{0.6\textwidth}}

\hline
\hline
0. CRPS estimates for each forecaster: & - For the $N$ couples forecast/observation, compute their    corresponding instantaneous CRPS.\\
  1. Estimation  of  $\gamma$ on the observations: & - Find a threshold $u$ where the Pareto approximation is acceptable and estimate the Pareto shape parameter $\gamma$ and $\sigma$ .\\
 \hline
 2. For a threshold $w \geq u$: & - Compute  the  scale parameter $\sigma_w= \sigma + \gamma w$.\\
   \hline
 3. Computation of   $X_u$&- Order the $m$ CRPS values where the observation $y \geq w$ in increasing order $s_1,\dots ,s_m$.\\
  For $i \in [1,m]$& -Compute for each CRPS value $s_i$, $H_{\gamma,\sigma_w}(s_i)$. \\
  & -Compute $\left[\frac{2i-1}{2m}-H_{\gamma,\sigma_w}(s_i)\right]^2$.\\ 
End 3.& \\
End 2. & \\
\hline
\end{tabular}
\end{center}
\end{table*}
Note that for large $u$, under the null hypothesis, 
the statistic ${\Omega^F_u}$ follows a Cram\'er-von Mises distribution.
The associated $p$-values $p^F_u \in [0,1]$ could have been computed, but they are actually subject to numerical instabilities \citep{prokhorov1968extension,csorgHo1996exact}. Furthermore, ${\Omega^F_u}$ is sufficient to compare the effect size of the deviation.




%
%
%

\end{document}